\documentclass[journal]{IEEEtran}
\usepackage{helvet}
\IEEEoverridecommandlockouts
\usepackage{paralist}
\usepackage{cite}
\usepackage{amsmath,amssymb,amsfonts}
\usepackage{graphicx}
\usepackage{textcomp}
\usepackage{amsmath}
\usepackage{xcolor}
\usepackage{graphicx} 
\usepackage{url}
\usepackage{algorithm}
\usepackage{array}
\usepackage{booktabs}
\usepackage{hhline}
\usepackage{algpseudocode}
\graphicspath{ {./fig/} }
\usepackage{mathtools}
\usepackage{multirow}
\usepackage{enumitem}
\setitemize{leftmargin=*, align=left}

\usepackage{amssymb}
\usepackage{orcidlink}

\graphicspath{ {./figures/} }
\def\BibTeX{{\rm B\kern-.05em{\sc i\kern-.025em b}\kern-.08em
    T\kern-.1667em\lower.7ex\hbox{E}\kern-.125emX}}
\begin{document}

\title{Signal Classification Recovery Across Domains Using Unsupervised Domain Adaptation}
\author{Mohammad~Ali\,$^{\orcidlink{0009-0007-6915-5261}}$,~\IEEEmembership{Student Member,~IEEE,}
	Fuhao~Li\,$^{\orcidlink{0000-0001-8076-2221}}$,~\IEEEmembership{Member,~IEEE,}
	and~Jielun~Zhang\,$^{\orcidlink{0000-0002-2113-2104}}$,~\IEEEmembership{Member,~IEEE}
	\IEEEcompsocitemizethanks{\IEEEcompsocthanksitem Mohammad Ali, Fuhao Li, and Jielun Zhang are with the School of Electrical Engineering and Computer Science, University of North Dakota, Grand Forks, ND, 58202. (email: \{ash.ali, fuhao.li, jielun.zhang\}@UND.edu)\protect\\
		}
\thanks{Manuscript received xxx; revised xxx; accepted xxx. }}

\markboth{Journal of \LaTeX\ Class Files,~Vol.~18, No.~9, September~2020}%
{How to Use the IEEEtran \LaTeX \ Templates}

\maketitle
\begin{abstract}
Signal classification models based on deep neural networks are typically trained on datasets collected under controlled conditions, either simulated or over-the-air (OTA), which are constrained to specific channel environments with limited variability, such as fixed signal-to-noise ratio (SNR) levels. As a result, these models often fail to generalize when deployed in real-world scenarios where the feature distribution significantly differs from the training domain. This paper explores unsupervised domain adaptation techniques to bridge the generalization gap between mismatched domains. Specifically, we investigate adaptation methods based on adversarial learning, statistical distance alignment, and stochastic modeling to align representations between simulated and OTA signal domains. To emulate OTA characteristics, we deliberately generate modulated signals subjected to realistic channel impairments without demodulation. We evaluate classification performance under three scenarios, i.e., cross-SNR, SNR-matched cross-domain, and stepwise adaptation involving both SNR and domain shifts. Experimental results show that unsupervised domain adaptation methods, particularly stochastic classifier (STAR) and joint adaptive networks (JAN), enable consistent and substantial performance gains over baseline models, which highlight their promise for real-world deployment in wireless systems.
\end{abstract}


\section{Introduction}

\IEEEPARstart{M}{odern} wireless communication systems demand robust and accurate signal classification in a wide variety of channel conditions, interference levels, and hardware environments. From cognitive radios to spectrum monitoring and military communication systems, the ability to distinguish modulation schemes is fundamental for downstream tasks like demodulation, resource allocation, or interference mitigation.

Traditional methods of performing signal classification are grounded in mathematical derivations that assume specific channel parameters. For example, likelihood-based approaches offer high performance under known channel conditions by calculating probabilities for each candidate modulation~\cite{ramezani2013likelihood}. Distribution-based methods use metrics such as Kolmogorov-Smirnov distance to match empirical distributions, e.g., symbol mapping distributions, with theoretical ones, requiring a large number of samples and stable signal-to-noise ratio (SNR) levels~\cite{6364879,urriza2011computationally}. Meanwhile, feature-based approaches often leverage hand-crafted features and traditional machine learning techniques. They offer flexibility but still struggled when confront with diverse or highly dynamic channel conditions.

With the rapid advancement of artificial intelligence and the availability of large datasets, deep learning has emerged as the go-to approach for signal classification. Specifically, neural networks can learn to extract high-level features directly from raw in-phase (I) and quadrature (Q) samples instead of requiring the need for manually engineered feature extractions~\cite{o2018over, west2017deep}. Deep learning architecture adapted from image processing, such as VGG and ResNet, have demonstrated strong performance by taking raw IQ data as input and automatically inferring the relevant modulation schemes~\cite{krizhevsky2012imagenet}. Hybrid architectures that combine convolutional neural networks (CNNs) with recurrent components, such as long short-term memory (LSTM) networks, appear even more adept, as they capture both spatial features and temporal memory, thus better modeling time-varying signals~\cite{luo2024rlitnn}. Recent work has further refined this direction by splitting IQ data into multiple branches, e.g., time, frequency, and other domains, and then fusing the extracted features into one classification head~\cite{9106397}.

Despite these significant advances, domain mismatch remains a major barrier to real-world deployment of deep learning based signal classification solutions. Models trained on clean simulated signals often fail when confronted with the unpredictability of over-the-air (OTA) data~\cite{bousmalis2017unsupervised}. Real-world factors such as hardware imperfections, unmodeled fading, varying interference, and inaccurate synchronization introduce discrepancies, i.e., domain shifts~\cite{o2016radio}, that can make a well-trained deep learning model underperform on the target data. Early attempts to mitigate this gap focused on fine-tuning the model’s higher layers on a small amount of labeled OTA data, serving as an approach that helps but depends on acquiring carefully labeled real, world samples~\cite{o2018over}. More recently, adversarial learning and domain adaptation techniques, e.g., domain adversarial neural networks (DANN), have been used to automatically align feature representations between simulated and OTA data, without the need for extensive OTA labels~\cite{ali2024automatic}.
However, many existing strategies still rely on partial labeling or simplified assumptions, which do not fully address the complexities of severely mismatched SNR environments.

In this work, we target the problem of signal classification across domains using unsupervised domain adaptation. Specifically, we systematically examine different adaptation strategies, including adversarial, distance-based, and stochastic approaches, and explore their efficacy in bridging the gap between simulation and OTA data under mismatched SNR levels. Extensive evaluations are conducted using simulated signal data and the OTA signal data generated in real world.
Specifically, our contributions in this paper are summarized as follows: (i) Creation of simulated and over-the-air datasets by generating radio signals of BPSK, QPSK, 8APSK, 16QAM modulations at varying SNR levels from 2dB to 22dB in steps of 4 dB. Real hardware is used to transmit and receive signals over the air; (ii) Implement several domain adaptation techniques for signal classification built on ResNet-18 model, including DANN~\cite{ganin2016domain}, maximum classifier discrepancy (MCD)~\cite{saito2018maximum}, deep correlation alignment (CORAL)~\cite{sun2016deep}, stochastic classifier (STAR)~\cite{lu2020stochastic}, and joint adaptive networks (JAN)~\cite{long2017deep}; (iii) Validate unsupervised domain adaptation in multiple scenarios, i.e., separate cross-SNR testing for simulated and over-the-air datasets, SNR-matched environments from simulated to over-the-air, and stepwise statically defined simulated SNR with varying over-the-air scenario.

The rest of this paper is structured as follows. Section~\ref{sec1} introduces the system model and signal model with both the overall framework and signal generation introduced. Section~\ref{sec2} describes the proposed unsupervised domain adaptation methods alongside methods of isolating domain shifts. Section~\ref{sec3} presents the experiments and the discussion. Section~\ref{sec4} concludes this work.

\section{Related Work}
\subsection{Traditional Signal Classification}
Classical methods for signal classification typically fall into three distinct categories, i.e., likelihood-based, distribution-based, and feature-based approaches. Likelihood-based classifiers compare the probability of each candidate modulation given the observed signal, yielding strong accuracy if channel parameters are well characterized. Nonetheless, these methods can become prohibitively expensive due to special-function evaluations in the probability density function, prompting research into approximation strategies such as Fourier and Tikhonov~\cite{ramezani2013likelihood,6155695}. Distribution-based approaches match empirical distributions of received samples to theoretical ones (e.g., via Kolmogorov–Smirnov or Kuiper tests), requiring a sizable dataset and prior knowledge of channel effects~\cite{6364879,sun2016deep}. In practice, they remain susceptible to performance degradation when SNR or phase jitter varies significantly~\cite{urriza2011computationally}. Feature-based classifiers take a more flexible route by extracting hand-engineered descriptors, e.g, higher-order statistical features~\cite{837045}, and training standard machine learning models like support vector machines, k-nearest neighbors, or Bayesian methods.
Although each of these methods retains value, they prove less adaptable in modern wireless systems for several reasons. First, likelihood-based classifiers can grow computationally unwieldy when dealing with broad or time-varying channel conditions. Second, distribution-based methods demand large sample sizes and accurate channel information, which may be unavailable or costly to obtain. Finally, feature-based techniques rely on careful feature engineering that can fail to capture the complexity of higher-order modulations and domain variability. As a result, these traditional pipelines often underperform when confronted with the rapidly shifting conditions and hardware imperfections characteristic of real-world deployment.

\subsection{Deep Learning for Signal Classification}
With the advent of artificial intelligence, deep learning has largely eclipsed earlier methods for signal classification. Large neural networks, supported by GPUs and automated feature extraction, have already revolutionized computer vision and natural language processing, and are increasingly applied to time-varying RF signals~\cite{krizhevsky2012imagenet}. For example, O'Shea \emph{et al}.~\cite{o2018over} pioneered the use of 1D CNNs, e.g., VGG, ResNet, on raw IQ data, showing higher accuracy than previous approaches. Follow-up work incorporated LSTM modules, forming Convolutional Long Short-Term Memory (CLDNN) models to exploit both spatial and temporal features~\cite{west2017deep}. Recent designs use multi-domain representations—time, frequency, power spectral density—and fuse them via attention or transformer-encoders~\cite{9106397, luo2024rlitnn, shao2024iqformer}.
Despite these advances, a recurring challenge arises when models trained on simulated datasets confront over-the-air signals. Transfer learning with a small number of labeled OTA samples partially bridges this gap~\cite{o2018over}, while domain-adversarial training seeks to learn domain-invariant features directly~\cite{ali2024automatic}. More generally, domain adaptation aligns distributions between source and target, mitigating performance drops caused by hardware imperfections, interference, or unmodeled channel effects~\cite{zhou2022domain}.

\subsection{Domain Shift and Adaptation Techniques}
As aforementioned, domain mismatch between simulated and OTA data presents a core challenge for signal classification. For example, hardware imperfections and unmodeled channel conditions cause real signals to deviate significantly from training data, thereby degrading model performance~\cite{wang2021multi}. Traditional fine-tuning on a small set of labeled OTA samples partially mitigates this gap, yet collecting high-quality labeled data can be impractical~\cite{gulrajani2020search}. Recently, unsupervised domain adaptation methods have gained traction for aligning source and target distributions using only unlabeled target data. Adversarial approaches, such as DANN, leverage a gradient reversal layer to learn domain-invariant features, while MCD maximizes then minimizes classifier disagreements on unlabeled target samples. Distance-based techniques, e.g., Deep CORAL, match the covariance of source and target feature spaces, whereas JAN employ multi-layer maximum mean discrepancy. Lastly, stochastic classifiers like STAR introduce randomness to capture predictive uncertainty and enhance feature alignment. These methods aim to maintain high accuracy on the labeled source domain while improving generalization to unlabeled target signals. Despite promising progress in computer vision and natural language processing, their application to real-world wireless environments, especially for modulations under severe SNR discrepancies, remains an active area of research~\cite{hoffman2018cycada,peng2019moment,ge2023domain}.

\begin{figure*}[t]
    \centering
    \includegraphics[width=1\linewidth]{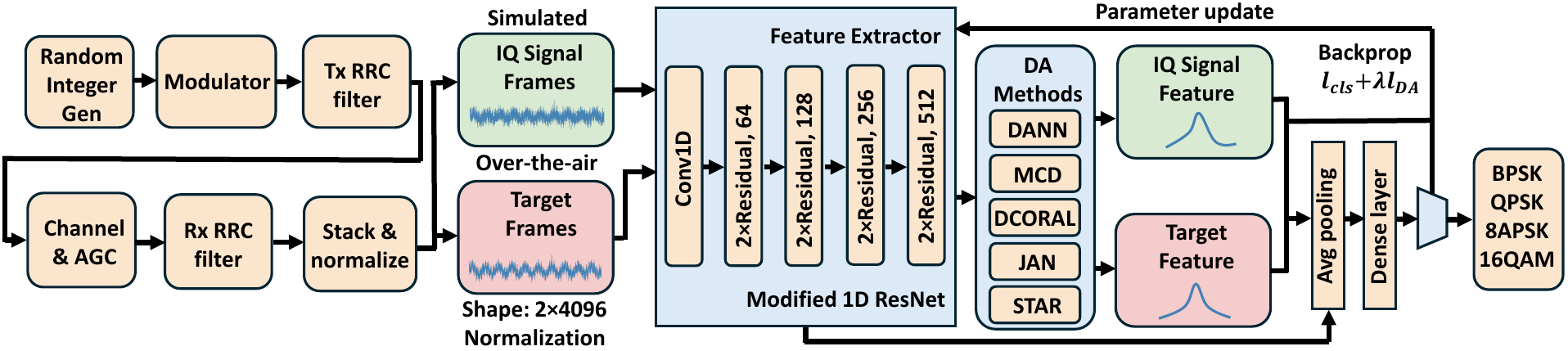}
    \caption{Overview of proposed framework for deep learning based signal classification using unsupervised domain adaptation. Domain shifts under consideration are shifts from simulated to OTA signals and SNR offsets.}
    \label{fig0}
\end{figure*}

\section{System and Signal Models}\label{sec1}

\subsection{System Model}
This works deep learning based signal classification method illustrated in Fig~\ref{fig0}, uses both simulated and over-the-air data to satisfy data domain shift. The simulated domain is limited in complexity, consisting of a single SISO fading channel with AWGN, and similar to over-the-air is constrained to SNR values between 2dB and 22dB. Limited synchronization and equalization in the simulated data domain results in an inherently cleaner received signal when compared to over-the-air which experience real-world channel characteristics not isolated or limited to conditions. Signal preprocessing includes a Root Raised Cosine filter and normalization when the complete signal is received to prevent bias in the feature extractor.
On the other hand, domain adaptation methods, including DANN, MCD, CORAL, JAN, and STAR, categorized into adversarial, distance, and stochastic based adaptation methods are each built on independent models because of differences in training method requiring varying networks to use both labeled source and unlabeled target data~\cite{zhou2022domain}. Specifically, a unified 1-D ResNet-18 architecture with identical layers, parameters and training configuration is the underlying model for each domain adaptation method focusing on comparison to the base model training only on labeled source data. 

\subsection{Signal Model}
In terms of signal modeling, data in a general form of random integers are fed to modulate signals, where each random integer is sampled uniformly from the same distribution using $m_k \sim \mathcal{U}\{0, 1, \dots, M-1\}$ where M is the total number of symbols present and ${m_k}$ are a sequence of random integers. The resulting independent and identically distributed random integers are modulated to obtain a transmit signal $s(t) = \sum_{k} d_k\, \delta(t - kT)$ where $d_k$ are the modulated symbols obtained from the mapped integers $m_k$. A channel model in relation to this transmit signal is given in the form
\begin{equation}
r(t) = \left( s(t) * h(t) \right) e^{j\left(2\pi \Delta f\, t + \phi\right)} + n(t),
\end{equation}
for which the received signal is defined by the transmit signal $s(t)$ convolved with the channel impulse response $h(t)$. Attenuation in a received signal is present in the form of frequency $\Delta f$, phase $\phi$ offsets and AWGN $n(t)$ as the noise channel. Inherent hardware synchronization offsets exist between a receiver and transmitter which are not accounted for, for simplicity. Simulated data as opposed to over-the-air differs as the attenuation present are only derived from a fading and AWGN channel. Signals are observed as IQ components and are organized into frames represented as
\begin{equation}
R^{(i)} = \begin{bmatrix}
\operatorname{Re}\{r^{(i)}[1]\},\operatorname{Re}\{r^{(i)}[2]\}, \dots,\operatorname{Re}\{r^{(i)}[N]\} \\
\operatorname{Im}\{r^{(i)}[1]\},\operatorname{Im}\{r^{(i)}[2]\},\dots,\operatorname{Im}\{r^{(i)}[N]\}
\end{bmatrix},
\end{equation}
which is a $2 \times N$ matrix for the $i$-th frame where each column is a single IQ quantized point and first row represents the real and second row represents the imaginary parts of a signal. Limiting each frame to $2 \times 4,096$ a normalized frame thus is defined by finding the maximum absolute value in an $i$-th frame by
\begin{equation}
M_i = \max_{m \in \{1,2\},\, n \in \{1,\dots,N\}} \left| R^{(i)}(m,n) \right|.
\end{equation}
Normalized frames provides stability during training in the neural network by ensuring a consistent range of values preventing bias towards certain signal amplitudes and mitigating outlier impact on model performance. The final normalized frames are processed into
\begin{equation}
\tilde{R}^{(i)}(m,n) =
\begin{cases}
\dfrac{R^{(i)}(m,n)}{M_i}, & \text{if } M_i \neq 0, \\
R^{(i)}(m,n), & \text{if } M_i = 0,
\end{cases}
\end{equation}
where $\tilde{R}^{(i)}(m,n)$ is the entire normalized frame in which each element is divided by the maximum absolute value $M_i$ if it is non-zero ultimately normalized frames are stacked in order of their SNR level, determined by the AWGN channel, and modulation type.

\section{Methodology}\label{sec2}
In this section, we present the methodological framework of this study. Specifically, we first model domain shifts specific to the wireless signal domain and present how these shifts are structured. Then, we introduce the unsupervised domain adaption strategies deployed in this work and formally describe their respective adaptation objectives. 

\subsection{Domain Shift}\label{seciva}
In this work, domains are characterized by either simulated or real-world signals and their respective SNR levels. The configurations employed when evaluating these domain shifts are illustrated in Fig.~2. Specifically, a signal to be classified using a neural network has a feature representation denoted by $X$ and is fed into a model to predict its modulation class $Y$; thus, a trained classifier is formally represented as $f: X \rightarrow Y$.

A domain shift is defined as adapting a model trained on a \emph{source domain} to perform accurately on a distinct \emph{target domain}. As an example, consider the domain difference arising from a high-SNR environment shifting toward a lower one, which can be quantitatively expressed as:  
\begin{equation}
    \delta_{\gamma_s,\gamma_t} = \|P_{\gamma_s}(X) - P_{\gamma_t}(X)\|,
\end{equation}
where the SNR levels $\gamma$ characterize specific domains $D_{\gamma}$, and $\delta_{\gamma_s,\gamma_t}$ denotes the difference in feature distributions between the source domain $P_{\gamma_s}(X)$ and the target domain $P_{\gamma_t}(X)$.

\begin{figure}[ht]
    \centering
    \includegraphics[width=\linewidth]{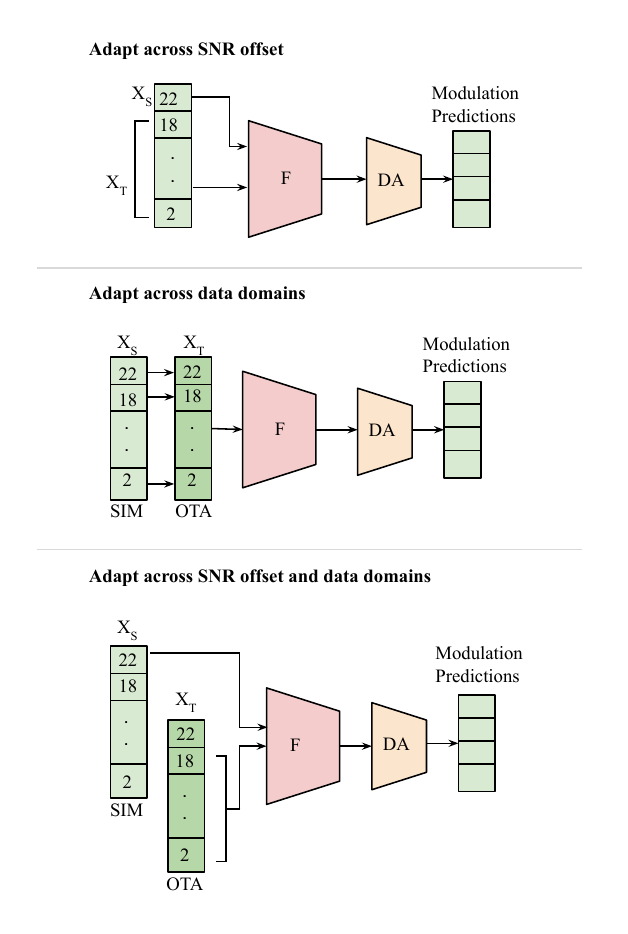}
    \caption{Overview of the studied domain shift methods. We evaluate on three domain shift possibilities, isolating the SNR offsets within a data domain, matching SNR offset and adapting cross-domain from simulated to over-the-air and finally stepwise SNR adaptation from simulated to over-the-air with no isolated domains.}
    \label{fig1}
\end{figure}

Due to the presence of such domain discrepancies, the performance of a model trained on the source domain $P_{\gamma_s}(X)$ typically deteriorates significantly when directly applied to the target domain $P_{\gamma_t}(X)$. To mitigate this degradation, our goal is to minimize the performance loss across domains, which can be mathematically formulated as:
\begin{equation}
    \min_{f \in \mathcal{F}} \mathcal{L}(f; D_{\gamma_s}) + \lambda \, \delta\left(P_{\gamma_s}(X), P_{\gamma_t}(X)\right),
\end{equation}
where the supervised classification loss on the source domain and the discrepancy between source and target domain distributions are balanced by a hyperparameter $\lambda$.
Three specific domain shift scenarios studied in this work (illustrated in Fig.~\ref{fig1}) are formulated as follows.

\subsubsection{{Cross-SNR Adaptation within the Same Domain}}
The first type of domain shift, easily quantifiable in signal classification, is a noise-level shift. Observing the performance degradation resulting from such a shift, the source domain SNR is fixed at the highest possible SNR level $D_{\Gamma}$, and the target domain varies across a range of lower SNR values $\gamma$. This scenario is mathematically described as:
\begin{equation}
    f_{\Gamma \rightarrow \gamma} = \arg\min_{f \in \mathcal{F}} \mathcal{L}(f; D_{\Gamma}) + \lambda \, \delta\left(P_{\Gamma}(X), P_{\gamma}(X)\right),
\end{equation}
which isolates the effects of varying noise conditions by training under optimal conditions and evaluating under degraded conditions.

 \subsubsection{SNR-matched Cross-domain Adaptation (Simulation to OTA)}
Another domain shift arises due to differences in the data domain (simulated vs. over-the-air signals), maintaining a fixed SNR level across the two domains. Formally, with source domain $D_{\gamma}^{\mathrm{sim}}$ representing simulated signals and target domain $D_{\gamma}^{\mathrm{ota}}$ denoting OTA signals, the scenario can be expressed as:
\begin{equation}
        f_{\gamma \rightarrow \gamma} = \arg\min_{f \in \mathcal{F}} \mathcal{L}(f; D_{\gamma}^{\mathrm{sim}}) + \lambda \, \delta\left(P_{\gamma}^{\mathrm{sim}}(X), P_{\gamma}^{\mathrm{ota}}(X)\right).
\end{equation}
This matched-SNR setting explicitly investigates domain discrepancies arising solely from inherent differences between simulated and realistic OTA signal characteristics.

 \subsubsection{Stepwise Domain Adaptation (Combined SNR and Data Domain)}
Finally, the most comprehensive scenario reflects simultaneous changes in both data type (simulated vs. OTA) and SNR levels, referred to as a true domain adaptation. In this case, no single domain-shift factor is isolated; rather, the classifier trained on simulated data with a fixed SNR level $\alpha$ must generalize robustly across various OTA SNR levels $\beta$. Mathematically, this scenario is defined as:
\begin{equation}
    f_{\alpha \rightarrow \beta} = \arg\min_{f \in \mathcal{F}} \mathcal{L}(f; D_{\alpha}^{\mathrm{sim}}) + \lambda \, \delta\left(P_{\alpha}^{\mathrm{sim}}(X), P_{\beta}^{\mathrm{ota}}(X)\right),
\end{equation}
which captures realistic challenges encountered by practical wireless signal classifiers, demanding robustness to simultaneous changes in channel characteristics and noise levels between source and target domains.

\subsection{Unsupervised Domain Adaptation for Signal Classification}
To explicitly address the domain shifts defined above, we adopt unsupervised domain adaptation techniques that align the distributions of IQ-frame representations from labeled source domains to unlabeled target domains. We denote the model architecture as composed of two primary components, i.e., a feature extractor $G_f: \mathbb{R}^{2\times4096} \rightarrow \mathbb{R}^{d}$ and a classifier $G_y: \mathbb{R}^{d} \rightarrow \mathbb{R}^{K}$. The goal is to minimize the combined objective that simultaneously ensures accurate classification on labeled source-domain samples and effective alignment of feature distributions between the source and target domains:
\begin{equation}
    \mathcal{L}_{\text{total}} = \mathcal{L}_{y} + \lambda\,\mathcal{L}_{\text{align}}(\mathcal{D}_S,\mathcal{D}_T),
\end{equation}
where $\lambda > 0$ is a hyperparameter balancing the supervised source-domain loss $\mathcal{L}_{y}$ and the unsupervised domain alignment loss $\mathcal{L}_{\text{align}}$. Specifically, the supervised classification loss on the source domain is defined using categorical cross-entropy as follows:
\begin{equation}
    \mathcal{L}_{y} = \mathbb{E}_{(x^s,y^s)\sim\mathcal{D}_S}\left[-\sum_{k=1}^{K}y_k^s\log\hat{y}_k^s\right],
\end{equation}
where $y^s$ is the ground truth represented in a one-hot encoding, and $\hat{y}^s = G_y(G_f(x^s))$ is the predicted class distribution.

\subsection{Implemented Instantiations for Unsupervised Domain Adaptation}
In this study, we systematically evaluate five representative unsupervised domain adaptation methods tailored specifically to the IQ signal classification task, each instantiating the alignment loss $\mathcal{L}_{\text{align}}$ differently:
\subsubsection{Adversarial Alignment (DANN)}  
This method introduces a domain discriminator $G_d:\mathbb{R}^{d}\rightarrow[0,1]$, which attempts to classify the domain origin (source or target) of feature representations. The feature extractor $G_f$ is simultaneously trained via a gradient reversal layer to deceive the discriminator, producing domain-invariant features:
\begin{align}
\mathcal{L}_{\text{align}}^{\text{DANN}} =\ & \mathbb{E}_{x\sim\mathcal{D}_S\cup\mathcal{D}_T}\big[ 
d\log G_d(G_f(x)) \nonumber \\
& + (1-d)\log(1-G_d(G_f(x))) \big]
\end{align}
where the domain label $d=1$ corresponds to source-domain samples and $d=0$ to target-domain samples. Thus, the DANN approach explicitly reduces the distributional gap identified in Section~\ref{seciva} by adversarially removing domain-specific patterns, such as SNR or OTA-induced impairments, while preserving modulation-specific characteristics.

\subsubsection{Maximum Classifier Discrepancy (MCD)}  
MCD leverages two parallel classifiers $G_{y_1}$ and $G_{y_2}$. The classifiers first intentionally maximize their disagreement on target-domain predictions, thus highlighting ambiguous and domain-specific features. Subsequently, the feature extractor $G_f$ is optimized to minimize this disagreement, enforcing alignment by moving target features closer to the source domain:
\begin{equation}
    \mathcal{L}_{\text{align}}^{\text{MCD}} = \mathbb{E}_{x^t\sim\mathcal{D}_T}\|G_{y_1}(G_f(x^t))-G_{y_2}(G_f(x^t))\|_{1}.
\end{equation}

\subsubsection{Deep CORrelation ALignment (Deep-CORAL)}  
Deep-CORAL explicitly matches the second-order statistics, i.e., covariance matrices, between source and target feature distributions. This statistical alignment directly targets SNR, induced amplitude scaling and OTA, induced variations:
\begin{equation}
    \mathcal{L}_{\text{align}}^{\text{CORAL}} = \|\text{Cov}(G_f(X_S))-\text{Cov}(G_f(X_T))\|_F^{2},
\end{equation}
where $X_S$ and $X_T$ represent mini-batches from source and target domains, respectively.

\subsubsection{Stochastic Classifier Alignment (STAR)}  
The STAR approach introduces model uncertainty by stochastically sampling classifier parameters from a Gaussian distribution. Alignment is enforced via prediction consistency measured using KL-divergence among multiple stochastic predictions on the target domain:
\begin{equation}
    \mathcal{L}_{\text{align}}^{\text{STAR}}=\sum_{m<n}\text{KL}(G_{y}^{(m)}(G_f(x^t))\|G_{y}^{(n)}(G_f(x^t))).
\end{equation}
This uncertainty-aware approach explicitly addresses unpredictable OTA distortions and channel effects not captured in simulation, thus stabilizing decision boundaries across domain shifts.

\subsubsection{Joint Adaptation Networks (JAN)}  
JAN extends the statistical alignment concept by applying Joint Maximum Mean Discrepancy (JMMD) simultaneously across multiple feature extraction layers. This multi-layer alignment captures higher-order and hierarchical interactions between temporal and spectral features, significantly relevant for recognizing modulation patterns affected by complex domain shifts:
\begin{equation}
    \mathcal{L}_{\text{align}}^{\text{JAN}} = \text{JMMD}\left(\{G_f^{\ell}(X_S)\}, \{G_f^{\ell}(X_T)\}\right),
\end{equation}
where $G_f^{\ell}$ denotes layer-specific activations. JAN thus effectively aligns deeper semantic representations that characterize modulation features across domains.

\section{Evaluation}\label{sec3}
In this section, we introduce the evaluation platform, baseline model, experimental datasets, and experimental scenarios. Specifically, the proposed domain adaptation models with 1D-customized ResNet-18 model is used for signal classification and rigorously tested in many configurations using simulated and over-the-air signals including cross-SNR for both datasets, SNR-matched cross-domain, and stepwise statically defined for true domain adaptation with no isolated domain shifts.

\subsection{Experimental Platform and Baseline Model}
\subsubsection{Wireless system parameters}
Signal generation was performed on two separate USRP NI 2901 boards configured with a center frequency of 916M MHz. A baseband sample rate of 1 MHz is used, and the filter employed is the root raised cosine filter with a rolloff factor of 0.35 and 8 samples per symbol for the receiver and transmitter. The USRP boards were positioned next to each other with a direct line of sight (LOS) for OTA signals. SNR is controlled using an AWGN channel and due to imperfections in OTA signals predefined SNR values set in the AWGN channel is used as the label for the received signals. In addition, the simulated signal generation differs only in its inclusion of a Rician fading model with a K-factor of 6 configured for LOS environment and no Doppler shift. Automatic gain control is enabled on the receiver end with a window length of 200 samples and a maximum gain of 60 dB to maintain a consistent signal level, providing reliable signals with no bias within the deep learning model.

\subsubsection{ResNet architecture}
\begin{figure}[ht]
    \centering
    \includegraphics[width=0.95\linewidth]{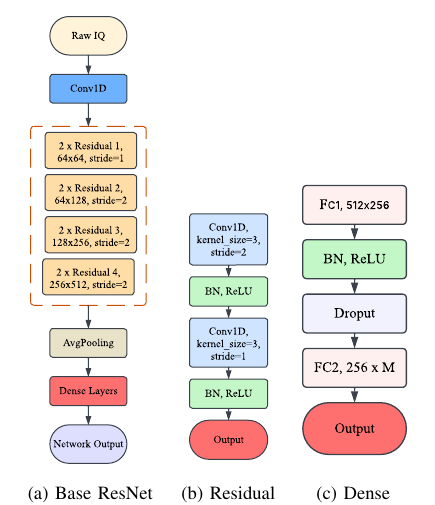}
    \caption{Base ResNet architecture, layers of residual and Blocks. Classifiers of DANN, CORAL, MCD, STAR, JAN use a dense block for end signal classification.}
    \label{fig2}
\end{figure}

The base model architecture for all domain adaptation methods employed in this work is a ResNet-18 illustrated in Fig.~\ref{fig2} adapted to process IQ signals. Specifically, the network layers begin with a 2x64 1D-convolutional layer, followed by four 2 times residual blocks, a global average pooling to obtain a fixed length of 512 feature dimension, which is passed through fully connected dense layers with dropout of 0.5 for improved training stability. Convolutional layers capture local feature patterns for short-term trends within each frame, while residual blocks aid in stable training, building a model that captures the inherent signal characteristics present through ordered IQ quantized points. The deeper layers capture higher-order abstractions related to the modulation types and channel-induced distortions.

Detailed representations of the residual blocks and dense layer in the ResNet are illustrated in Fig.~\ref{fig2}(b) and Fig.~\ref{fig2}(c), respectively. The architecture used in this work contains four residuals, each containing two blocks, with features increasing from 64 to 512 features. Each residual block includes two separate 1D convolution layers, both with batch normalization and ReLU activation. Each Convolution has a kernel size of 3 and a stride of 2. In addition to the residual blocks, the ResNet architecture contains dense layers used for classification. A separate classifier network uses this dense layer for each adaptation method, serving as the separation between feature extraction and classification, but it is not applied to the baseline model where a single end-to-end architecture is used.

\subsubsection{Compute environment}
All models execute on NVIDIA RTX 4090 using the same set of parameters for training the models and configuration within each adaptation method for reproducibility. The parameters used in the model training phases consist of a learning rate of 0.001, 50 epochs, 5 runs per test for averaging out the results, batch size of 128, Adam optimizer, and cross-entropy for classifying the different modulations.

\subsection{Experimental Datasets}
The dataset is introduced in Table~\ref{tab0}, totaling 24,576 samples over four modulations, i.e., BPSK, QPSK, 8APSK, 16QAM. Each modulation is sampled in the range of 2dB to 22dB, including steps of 4dB, and each SNR level contains 1,024 samples for every modulation. A resulting frame is in size $2 \times 4,096$. Thus, a model has to compute a total of 50,331,648 IQ points. In the simulated dataset, Rician fading and AWGN channels collectively simulate the channel characteristics, whereas in over-the-air, the real-world channel conditions do so instead. Models are trained and tested on a balanced number of samples across all modulation types, and results are averaged through 5-fold cross-validation for reproducibility and to ensure an accurate set of results throughout the training and inference phases.
\begin{table}[ht]
\caption{Summary of experimental datasets.}
\label{tab0}
\centering
\begin{tabular}{|c|>{\centering\arraybackslash}p{3cm}|}
\hline
\textbf{Element} & \textbf{Value} \\ \hline
\emph{Sample size} & 24,576 \\ \hline
\emph{Frame shape} & IQ: 2 x 4,096 \\ \hline
\emph{Total quant} & 50,331,648 \\ \hline
\emph{Modulations} & BPSK, QPSK, 8APSK, 16QAM \\ \hline
\emph{SNR range} & 2dB $\sim$ 22dB \\ \hline
\emph{SNR step} & 4dB \\ \hline
\emph{Samples per SNR} & 1,024 x 6 \\ \hline
\emph{Channel characteristics} & AWGN, Rician Fading Channel \\ \hline
\end{tabular}
\end{table}

\subsection{Experimental Analysis}
This paper evaluates three categories of experiments. The first is to isolate the SNR variance within a data domain and treat each SNR offset as an independent domain shift to verify the potential of domain adaptation across noise levels, referred to as cross-SNR. The second experiment focuses on data domains as the main shift by adapting models trained on simulated data to over-the-air with no SNR offset, thus called SNR-matched cross-domain. The third experiment represents a more realistic unsupervised domain adaptation scenario, where no specific domain shifts are isolated by freezing a model trained on a singular simulated SNR level against varying over-the-air noise to verify the ability of unsupervised domain adaptation on the real-world configuration. A comparison is made by freezing all possible simulated SNR offsets to verify the similarity of data domains as a determining factor in recovering performance across domains.

\begin{figure}[ht]
    \centering
    \includegraphics[width=\linewidth]{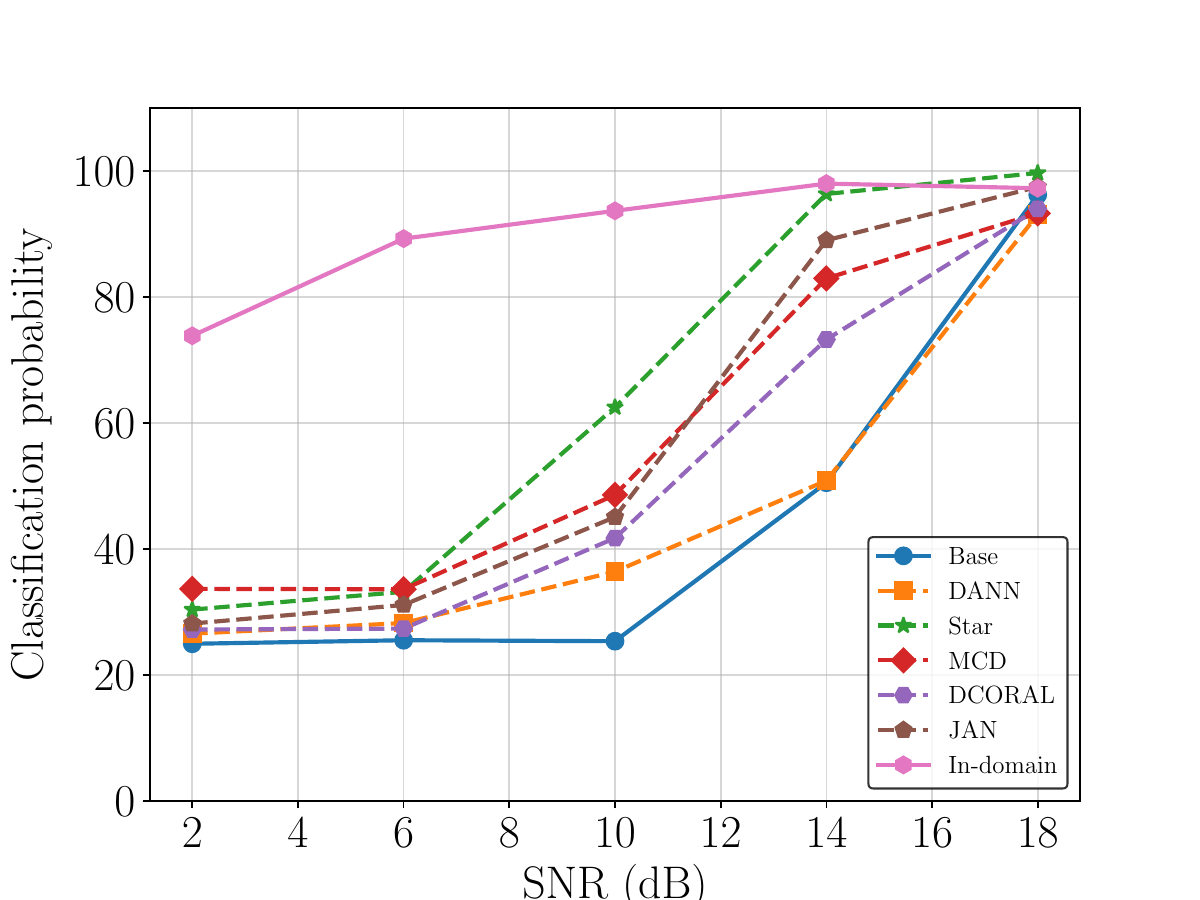}
    \caption{Cross-SNR classification probability for simulated signals only with source as 22dB.}
    \label{fig3}
\end{figure}

\begin{table*}[h]
\centering
\caption{Cross-SNR detailed results for simulated data.}
\label{tab1}
\setlength{\tabcolsep}{4pt}
\begin{tabular}{|c||ccc||ccc||ccc||ccc||ccc|}
\hline
\multirow{2}{*}{\textbf{Method}} & \multicolumn{3}{c||}{\textbf{2dB}} & \multicolumn{3}{c||}{\textbf{6dB}} & \multicolumn{3}{c||}{\textbf{10dB}} & \multicolumn{3}{c||}{\textbf{14dB}} & \multicolumn{3}{c|}{\textbf{18dB}} \\ \cline{2-16} 
& Mean & Max & Var & Mean & Max & Var & Mean & Max & Var & Mean & Max & Var & Mean & Max & Var \\ \hline
Base      &24.96&25.35&00.08   &25.52&27.47&00.96   &25.37&26.74&00.48   &50.52&54.98&14.26    &96.22&99.82&07.59   \\
In-domain &73.84&78.53&08.72   &89.26&91.12&00.97   &93.68&99.76&16.95   &98.00&100.00&06.88    &97.27&98.98&01.57   \\
DANN      &26.56&29.76&04.32   &28.25&33.09&09.47   &36.40&45.18&38.92   &50.85&65.73&121.68    &93.09&95.73&02.93   \\
CORAL     &27.21&28.64&00.99   &27.37&28.74&00.64   &41.74&45.49&04.84   &73.25&81.90&28.91    &93.98&99.15&33.31   \\
JAN       &28.16&29.88&00.80   &31.12&34.51&03.33   &45.07&50.38&21.51   &89.01&92.97&08.64    &97.48&99.83&15.62   \\
MCD       &33.67&38.66&10.48   &33.61&36.31&03.73   &48.59&55.82&16.09   &82.95&88.45&12.28    &93.29&99.10&20.16   \\
STAR      &30.38&31.53&01.29   &33.21&34.56&01.42   &62.50&66.03&05.18   &96.36&97.80&00.91    &99.68&99.83&00.01   \\ \hline
\end{tabular}
\end{table*}

\begin{figure}[ht]
    \centering
    \includegraphics[width=\linewidth]{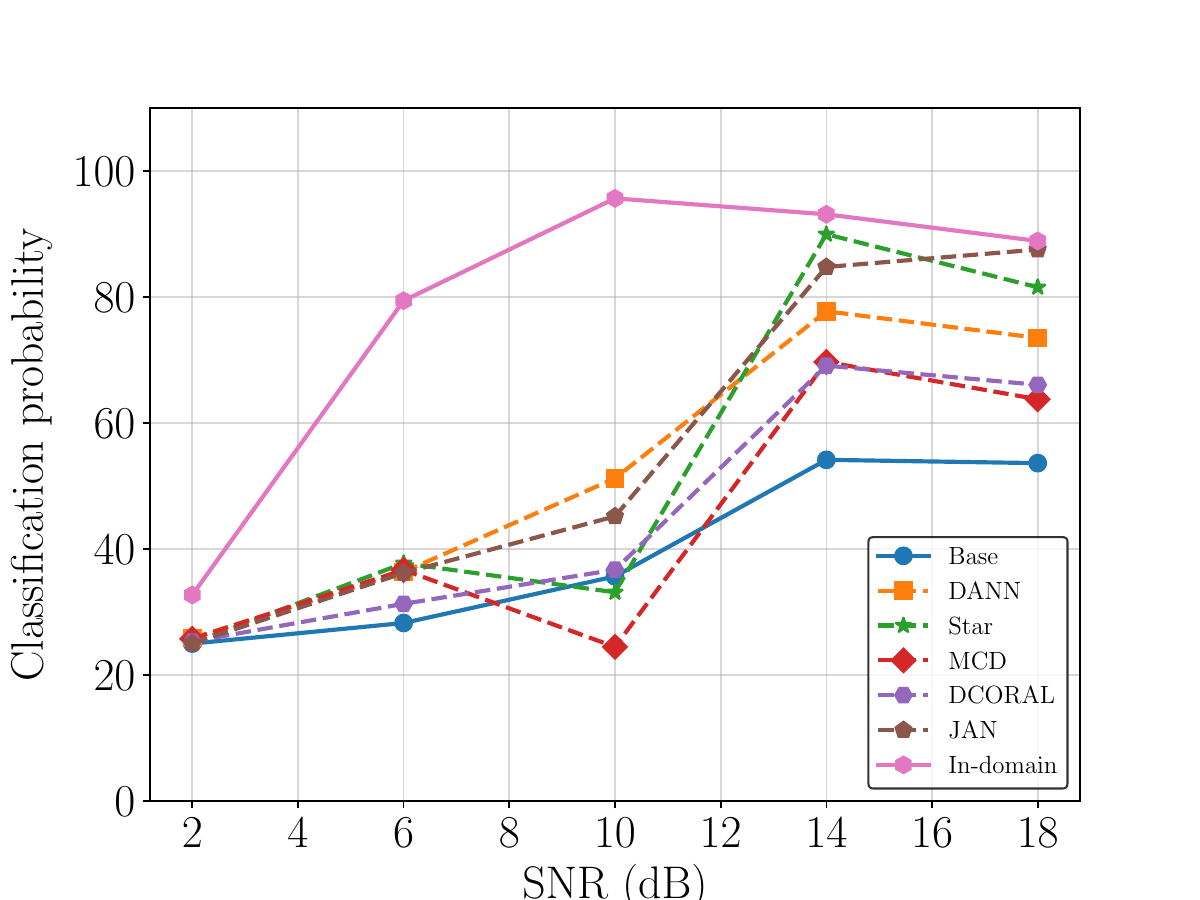}
    \caption{Cross-SNR classification probability for OTA signals only with source as 22dB.}
    \label{fig4}
\end{figure}

\begin{table*}[h]
\centering
\caption{Cross-SNR detailed results for over-the-air data.}
\label{tab2}
\setlength{\tabcolsep}{4pt}
\begin{tabular}{|c||ccc||ccc||ccc||ccc||ccc|}
\hline
\multirow{2}{*}{\textbf{Method}} & \multicolumn{3}{c||}{\textbf{2dB}} & \multicolumn{3}{c||}{\textbf{6dB}} & \multicolumn{3}{c|}{\textbf{10dB}} & \multicolumn{3}{c||}{\textbf{14dB}} & \multicolumn{3}{c|}{\textbf{18dB}} \\ \cline{2-16} 
& Mean & Max & Var & Mean & Max & Var & Mean & Max & Var & Mean & Max & Var & Mean & Max & Var \\ \hline
Base      &24.98&25.45&00.08   &28.26&30.94&02.50 &35.65&41.49&19.33  &54.15&60.37&13.01    &53.63&58.83&18.01   \\
In-domain &32.70&36.20&04.33   &79.41&92.59&56.35   &95.65&97.90&02.09   &93.13&99.66&17.21    &88.89&100.00&37.98   \\
DANN      &25.85&26.62&00.35   &36.35&47.26&55.95   &51.23&73.02&205.00   &77.73&87.18&58.56    &73.51&75.95&04.26   \\
CORAL     &25.29&25.83&00.21   &31.30&36.34&09.63   &36.71&39.15&07.25   &69.11&73.50&06.31    &66.06&71.43&09.40   \\
JAN       &25.01&25.45&00.05   &36.20&39.10&11.01   &45.19&49.99&11.36   &84.76&89.45&22.03    &87.54&98.56&33.43   \\
MCD       &25.73&27.16&00.79   &36.69&41.83&14.50   &24.46&31.72&21.28   &69.69&79.02&32.00    &63.77&73.76&32.01   \\
STAR      &25.05&25.61&00.12   &37.74&40.54&02.85   &33.13&37.85&15.26   &89.96&96.36&14.61    &81.55&94.51&46.85   \\ \hline
\end{tabular}
\end{table*}

\subsubsection{Cross-SNR}
Figure~\ref{fig3} illustrates the classification probability of a model trained exclusively on simulated data sourced at 22dB SNR and tested on a target domain ranging from 2dB to 18dB. A comparison of domain adaptation methods is made in relation to a baseline approach with no adaptation alongside the in-domain performance, which is a display of a model trained and deployed in the same domain as the target, indicating no domain shift presence. Besides DANN, all adaptation methods are considerably better than the baseline approach as shown in Table~\ref{tab1} with the highest recovery around 10dB to 14dB with probabilities of $46.86\%$ and $78.49\%$ averaged across all adaptation methods compared to $25.37\%$ and $50.52\%$ for baseline. Interestingly, we observe the STAR method outperforming the in-domain results with $99.68\%$ compared to $97.27\%$ at 18dB. At the same time, other adaptation curves rapidly converge closely towards the in-domain performance as SNR difference decreases between the source and target domains, averaging out to $95.50\%$ with $97.27\%$ for the in-domain. Ideally, an improvement in classification probability while converging source and target domains is expected. However, we note that near-perfect classification is a consequence of low complexity in features of simulated data.

A similar experiment is computed exclusively with over-the-air data illustrated in Fig~\ref{fig4}. Specifically, a difference is present in overall classification probability shown in Table~\ref{tab2} with in-domain averaging $77.96\%$ compared to $90.41\%$ for simulated data across all SNR levels due to the greater dynamic channel attenuation present in over-the-air signals. Other notable differences are observed between the data domains, i.e., the best recovery is present at 14dB to 18dB, whereas it peaks at 10dB to 14dB for simulated, believed to be due to different channel distortions besides noise and fading that affect real-world signals, which may not exist in simulated data. All adaptation methods are observed as improvements over baseline apart from STAR and MCD, which classify with a probability of $33.13\%$ and $24.46\%$ from a baseline of $35.65\%$. Note that the baseline classifier peaks at $54.15\%$ and never converges close to in-domain results as source and target differences are reduced. JAN is the only adaptation method that does come close with a probability of $87.54\%$ compared to $88.89\%$ while the averaged result is $74.49\%$ for all adaptation methods, and no single method has the highest classification probability across all SNR levels compared to simulated results where STAR does. However, neither the base nor any adaptation methods recover enough probability in the lower to mid SNR levels for both simulated and over-the-air.

\begin{figure}[ht]
    \centering
    \includegraphics[width=\linewidth]{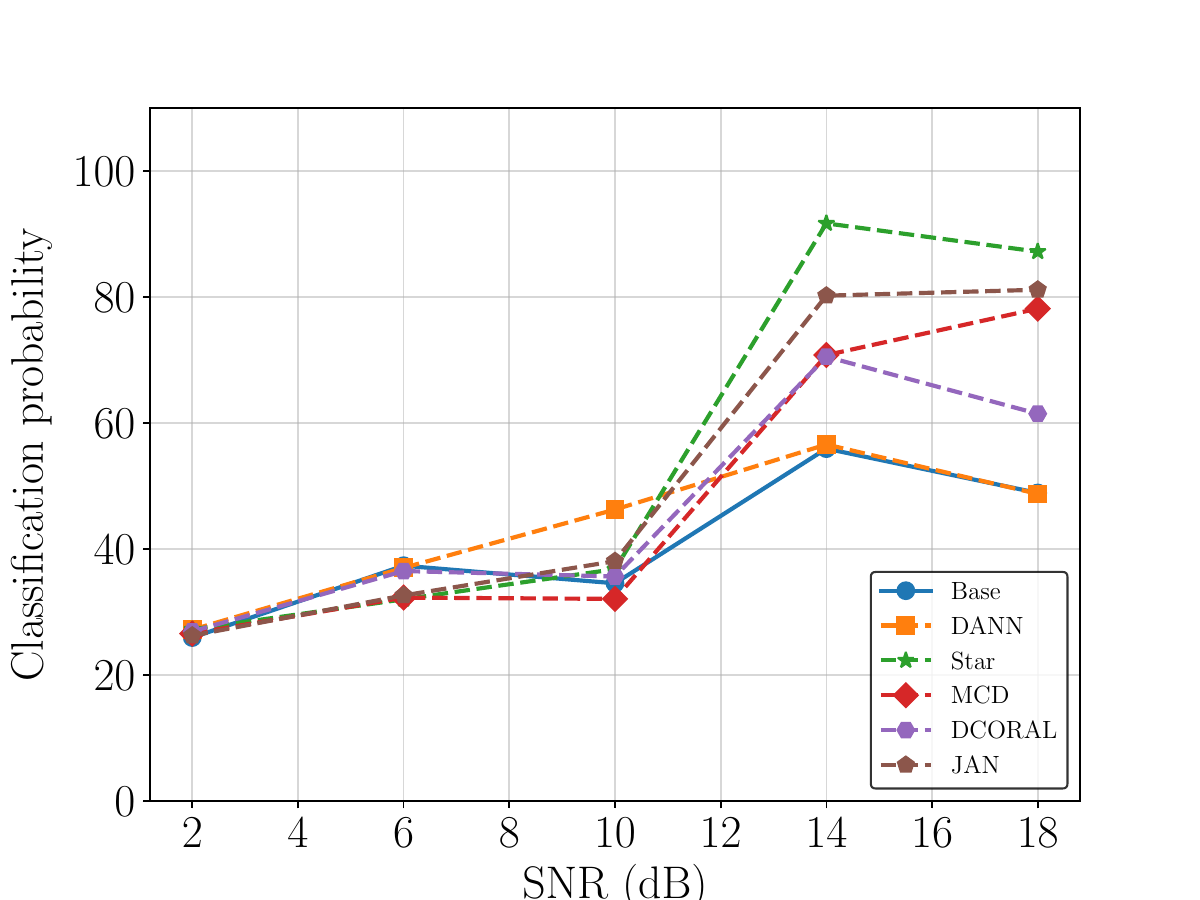}
    \caption{SNR-matched classification probability. Distribution shift in simulated to OTA while SNR is matched between source and target.}
    \label{fig5}
\end{figure}

\begin{figure*}[ht]
    \centering
    \includegraphics[width=\linewidth]{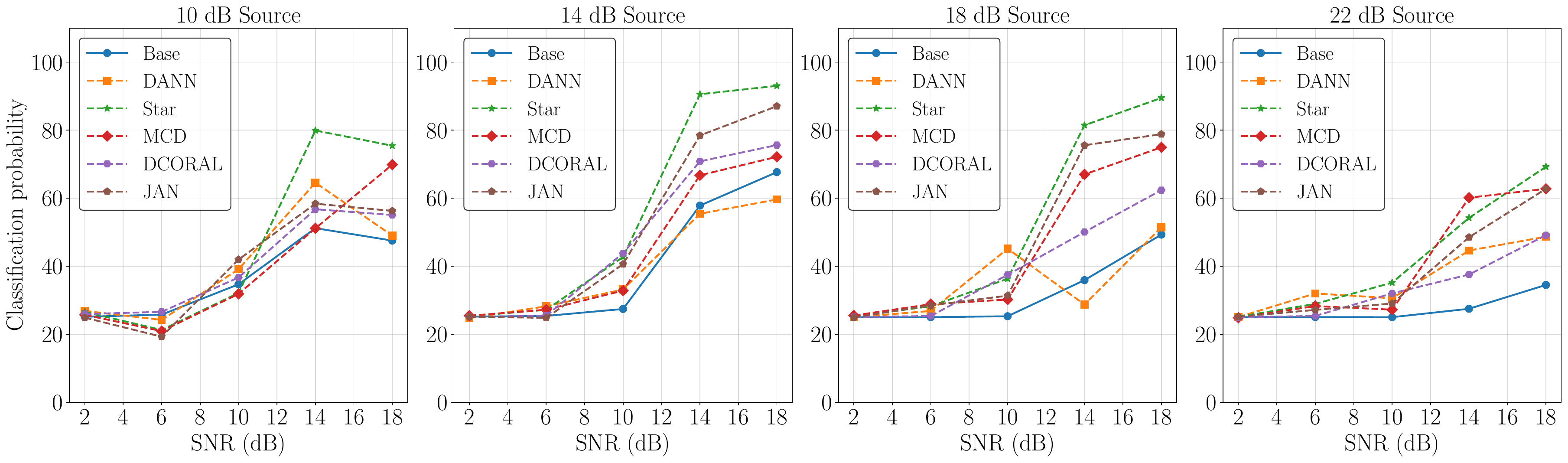}
    \caption{Evaluation result of model trained on simulated data adapted to OTA with source SNR selected in stepwise approach and varying target SNR. For simplicity plots at source SNR 2dB and 6dB are not included because of poor performance for all target SNR offsets.}
    \label{fig6}
\end{figure*}

\begin{table*}[ht]
\centering
\caption{SNR-matched adapting from simulated to over the air.}
\label{tab3}
\setlength{\tabcolsep}{4pt}
\begin{tabular}{|c||ccc||ccc||ccc||ccc||ccc|}
\hline
\multirow{2}{*}{\textbf{Method}} & \multicolumn{3}{c||}{\textbf{2dB}} & \multicolumn{3}{c||}{\textbf{6dB}} & \multicolumn{3}{c||}{\textbf{10dB}} & \multicolumn{3}{c||}{\textbf{14dB}} & \multicolumn{3}{c|}{\textbf{18dB}} \\ \cline{2-16} 
& Mean & Max & Var & Mean & Max & Var & Mean & Max & Var & Mean & Max & Var & Mean & Max & Var \\ \hline
Base   &25.94&26.39&00.12   &37.35&41.47&09.92   &34.55&37.73&04.80   &55.93&62.00&19.68    &48.90&52.07&08.02   \\
DANN   &27.20&29.55&01.90   &37.01&46.03&40.82   &46.26&59.10&104.76   &56.59&71.46&99.66    &48.75&76.83&307.29   \\
CORAL  &26.91&27.52&00.26   &36.50&38.17&00.85   &35.64&39.10&05.59   &70.52&75.12&23.76    &61.47&66.74&19.56   \\
JAN    &26.26&27.64&01.35   &32.63&36.26&06.96   &38.07&41.90&09.07   &80.21&85.99&20.52    &81.15&83.86&08.27   \\
MCD    &26.56&28.08&01.74   &32.29&37.56&08.60   &32.08&34.16&04.01   &70.78&73.94&04.20    &78.15&87.99&31.10   \\
STAR   &27.23&30.11&03.26   &32.02&33.72&01.46   &36.80&46.79&41.16   &91.69&93.33&02.17    &87.20&89.94&04.12   \\ \hline
\end{tabular}
\end{table*}

\subsubsection{SNR-matched}
The experiment results in the SNR-matched scenairo are illustrated in Fig~\ref{fig5} where the source data is simulated and adapted to the target containing over-the-air data in a configuration such that the SNR levels are equal from 2dB to 18dB. Similar to cross-SNR for over-the-air data, peak recovery is at 14dB and levels off at 18dB SNR. However, the base does perform better than other methods at the range of 2dB to 10dB and slightly better than DANN at 18dB, which suggests channel effects are different between simulated and over-the-air at these noise levels as shown in Table~\ref{tab3}. The greatest improvements achieved through domain adaptation exist at higher SNR levels, notably at 14dB, presenting a $35.76\%$ difference in base ($55.93\%$) and STAR ($91.69\%$) and even higher at 18dB, a $38.30\%$ improvement from base ($48.90\%$) to STAR ($87.20\%$). A glance at the averaged probability at these two SNR levels for all adaptation methods shows $73.96\%$ and $71.34\%$, which is an overall recovery difference of $18.02\%$ and $22.44\%$, respectively. Analyzing the adaptation models' stability reveals that variance rises as the SNR level does, even more so for DANN, where the maximum classification values achieved are $71.46\%$ and $76.83\%$. In contrast, the mean values are $56.59\%$ and $48.75\%$ at 14dB and 18dB, respectively. This indicates that DANN produces highly inconsistent classifications across runs at higher SNR levels, suggesting sensitivity in its feature-alignment process when dealing with cleaner signals as opposed to methods like STAR, JAN, and MCD, which generally show much lower variance across all SNR levels demonstrating greater consistency in adaptation runs. These results suggest adaptation from simulated to over-the-air is possible when noise levels are equal for shifts within acceptable means as opposed to the lower to mid-SNR range, where the model struggles for base and domain adaptation due to significant differences between the data domains.

\begin{figure*}[ht]
    \centering
    \includegraphics[width=0.88\linewidth]{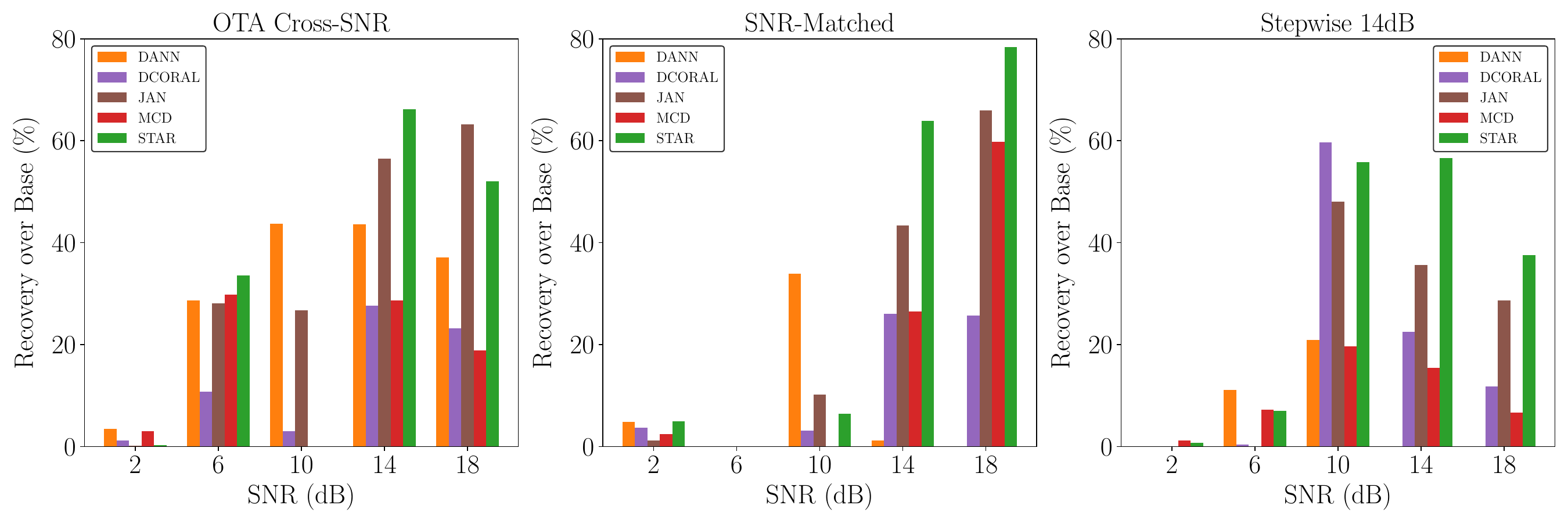}
    \caption{Domain adapted models recovery of classification probability over the base model in cross-SNR for OTA, SNR-matched and stepwise (selection of 14dB). Note that the methods that do not appear to have any recovered performance may have worse result than base model i.e. 6dB for SNR-matched.}
    \label{fig9}
\end{figure*}

\subsubsection{Stepwise}
A detailed set of results from a stepwise adaptation experiment, designed to closely resemble realistic scenarios, is illustrated in Fig~\ref{fig6}, where simultaneous shifts occur in the SNR and the data domain. Each subplot is a model trained on simulated source data from 10dB to 22dB and adapted to target OTA of varying SNR from 2dB to 18dB. When the simulated source domain is fixed at moderate SNR levels of 14dB and 18dB, the domain adaptation methods show consistent significant improvements over the baseline, especially at the higher end of the target SNR where cleaner noise conditions are present. The models trained and validated on moderately distorted simulated signals are most effective at adapting to real-world signals, and the greatest adaptation method among them is STAR, consistently achieving the highest classification probability, followed by JAN and MCD, showing steady improvements. The recovery gained by using domain adapted classifiers declines at both extremes of simulated SNR ranges. For example, when the source is set to 22dB, the classification probability curves show a flatter progression peaking around $60.00\%$ using STAR, JAN, and MCD methods, whereas the base tops at $34.49\%$. The results show diminishing returns using a clean source signal as opposed to over-the-air signals, which are not limited to any channel impairments besides AWGN and fading. Not shown in the figure are when the source is set to 2dB and 6dB, respectively, due to extremely poor performance with minimal to no improvement for all methods, including base, proving a significant challenge for any recovery when signals trained on are high in noise. This experiment shows an optimal unsupervised domain adaptation approach, which uses the intermediate SNR levels as opposed to both extreme ends, highlighting a balance in channel distortion to achieve the greatest domain adaptation results and carefully choosing which adaptation method to be used.

A summary of recovery gained by domain adaptation methods over the baseline in percent is illustrated in Fig.~\ref{fig9} for cross-SNR in OTA, SNR-matched, and stepwise with a selected source of 14dB. It can be seen in cross-SNR configuration recovery is present between 6dB and 18dB with negligible gains at 2dB, which suggests that the domain shifts are isolated within only SNR and other real-world attenuation. We can expect our method to have considerable gains for many SNR offsets. This is not the case when data shifts are present going from simulated to OTA, as SNR-matched gains are present at 14dB and 18dB only. However, the impact of recovery is even greater than in cross-SNR, especially for STAR and JAN methods. In the stepwise adaptation scenario, when both data type and SNR levels differ between the source and target domains, performance gains are observed primarily between 10dB and 18dB. Among the evaluated methods, CORAL achieved the highest peak performance, while STAR consistently ranked among the top-performing techniques. Notably, selecting a source domain with an SNR of 14dB yielded the most substantial recovery at target SNRs of 10dB and 14dB, which is likely due to their distributional similarity. As the gap between source and target SNR widens, the adaptation effectiveness declines. These results highlight the critical importance of selecting an appropriate adaptation strategy. STAR and JAN demonstrated reliable recovery across varying shifts, whereas DANN showed limited effectiveness in SNR-matched and stepwise scenarios. However, DANN still offered moderate gains in cross-SNR settings, suggesting its utility is confined to adaptation within the same data domain.

\section{Conclusion}\label{sec4}
This study explored unsupervised domain adaptation for deep-learning-based signal classification under domain shifts caused by varying SNR levels and real-world channel conditions. We implemented and evaluated five UDA methods, i.e., DANN, MCD, Deep-CORAL, JAN, and STAR, across three distinct scenarios, i.e., cross-SNR adaptation, SNR-matched cross-domain transfer, and stepwise adaptation from simulated to OTA signals. Experiment results demonstrate that unsupervised domain adaptation methods can significantly mitigate performance degradation from domain mismatch. Among them, STAR and JAN consistently performed well, particularly at moderate to high SNR levels, demonstrating robustness to real-world signal impairments. We observed that mid-range SNR levels in the source domain offer the best foundation for adaptation, while extremely low SNRs yield less transferable features. The effectiveness of each unsupervised domain adaptation method also varied depending on the nature of the domain shift. These findings provide practical guidance for deploying deep learning models in wireless systems and emphasize the need for proper selection of adaptation strategies and training conditions. In future, we aim to investigate online adaptation techniques for real-time signal streams and explore multi-source domain adaptation techniques to further improve the classifier robustness in deployment.

\renewcommand\refname{References}
\bibliographystyle{IEEEtran}
\bibliography{IEEEfull,Reference}

\end{document}